\documentclass[journal]{IEEEtran}

\pdfoutput=1

\usepackage{amsfonts}
\usepackage{cite}
\usepackage{amsmath}
\usepackage{graphicx}
\usepackage{epstopdf}
\usepackage{amsthm}
\usepackage{color}

\usepackage{verbatim}

\usepackage{array}

\usepackage{kbordermatrix}

\begin{document}

\title{Improved steady-state stability of power grids with a communication
infrastructure}

\author{A. Gajduk, M. Todorovski, and
        L. Kocarev,~\IEEEmembership{Fellow,~IEEE}%
\thanks{Copyright (c) 2012 IEEE. Personal use of this material is permitted. However, permission to use this material for any other purposes must be obtained from the IEEE by sending an email to pubs-permissions@ieee.org.}
\thanks{A. Gajduk is with the Macedonian Academy of Sciences and Arts, Skopje, Macedonia. E-mail: agajduk@manu.edu.mk}

\thanks{M. Todorovski is with the Faculty of electrical engineering and information technologies, University ``Ss Cyril and Methodius'', Skopje, Macedonia. Email: mirko@feit.ukim.edu.mk}

\thanks{L. Kocarev is with the Macedonian Academy of Sciences and Arts, Skopje, Macedonia; the Faculty of Computer Sciences and Engineering, University ``Ss Cyril and Methodius'', Skopje, Macedonia and the BioCircuits Institute, UC San Diego, La Jolla, CA 92093-0402, USA. E-mail: lkocarev@ucsd.edu}
}

\maketitle

\begin{abstract}
Efficient control of power systems is becoming increasingly difficult as they gain in complexity and size. 
We propose an automatic control strategy that regulates the mechanical power output of the generators in a power grid based on information obtained via a communication infrastructure.
 An algorithm that optimizes  steady-state stability of a power grid by iteratively adding communication links is presented. The proposed control scheme is successfully applied to the IEEE New England and IEEE RTS 96 power systems, leading to a significant increase in the steady-state stability of the systems and an improvement in their overall robustness. The resulting communication network topology differs significantly from the transmission grid topology. This shows how complex the steady- state control for power systems is, influenced by the generators configuration, the transmission network topology, and the manner by which control is executed.
\end{abstract}


\section{Introduction}
\label{intro}

Steady-state stability is the study of the power system stability when the system is subjected to small perturbations such as incremental changes in system load.
These disturbances are considered to be sufficiently small that linearization of the system equations is permissible for purposes of analysis.
There is extensive literature concerning linearized models in power systems \cite{kundur1990comprehensive,sauer1990power,wang1989application}.  They are obtained numerically from the nonlinear differential algebraic models. Linear models are not only useful for  small signal stability analysis of power systems, but also for designing power systems control techniques.

Presently, available techniques for linearization  may be categorized mainly into three types \cite{yadaiah2007linearisation} : (1) standard form of linearization technique using Taylor series approximation, (2) direct feedback linearization and (3) linearization of nonlinear model based on differential geometric technique.
Recent studies address steady-state stability in power systems which operating conditions are facing increased uncertainty due to the integration of renewable resources such as wind energy conversion systems and photovoltaic energy conversion systems, and loads such as plug-in hybrid electric vehicles \cite{dominguez2012reliability}.

Electrical energy today is produced by large synchronous generators driven by a prime mover, usually a turbine~\cite{wood2012power}.
When the total generation is equal to the total system demand the frequency is constant and the system is in equilibrium. 
If this state is to be maintained a change in the delivered electrical power must be compensated by an equal change in the mechanical power of the generators turbines.
Any changes in the system load are initially supplied by grid inertia, which represents the total combined kinetic energy stored in a power system found in the large rotating masses of the generators and turbines.
However, grid inertia can be quickly spent if additional action is not taken.

Mechanical power adjustment in the generators is performed by a separate module, called the \textit{turbine governor}. The turbine governor aims to maintain the frequency of the system at a scheduled value. This is also referred to as frequency control.
Before the application of real-time phasor measurements, control was essentially local, with actions taken based on local measurements sometimes supported by mathematical models for the rest of the system~\cite{moussa1974dynamic,cheng1986self,kundur1989application}.

Phasor measurement units (PMU) or synchrophasors can measure the electrical waves on an electricity grid by employing a common time source for synchronization purposes. This enables monitoring of multiple remote points on the gird in real-time and a whole new set of control possibilities.
Synchrophasors are already being applied to various control problems, such as excitation control, power system stabilizers and FACTS devices control~\cite{zhang2008design,
heniche2008assessment,de2010synchronized,qiao2007dhp,hassan2009power}.

The majority of authors employ linear feedback control system and show that they are capable of coordinating a number of local control units. These local controllers have inputs from a variety of monitoring devices placed on numerous locations around the grid.
A measurement unit can be placed anywhere on the grid.
Selecting the optimal locations is thus an important decision as it affects the overall performance of the control scheme.

We propose a control strategy for improving the steady-state stability of the power systems, in which the mechanical power output of each generator is governed using information from PMUs placed at various locations around the grid, in addition to its local operational conditions.
A similar idea on this topic is proposed in~\cite{li2011synchronization}. The authors in~\cite{li2011synchronization} also exhibit control by using information gathered from remote locations on the grid which are transmitted using a communication infrastructure. However, their study is limited to power grids that consist of identical generators, thus making the entire proposal hard to test on existing systems and to apply in reality.

The outline of the paper is as follows. The principal idea along with an algorithm for constructing the communication infrastructure are presented in Section~\ref{steady_state}. This section also describes a power system model adopted for stability analysis. The results of applying the control scheme to the New England and IEEE RTS 96 power systems are presented in Section~\ref{results}. Section~\ref{concl} concludes this paper.

\section{Steady-state stability analysis}
\label{steady_state}

Steady-state stability has always played an important role in studying, understanding and designing power systems.
A power system is said to be steady-state stable if it is able to reach a new stable configuration following a small disturbance in the system. Load fluctuation, actions of automatic voltage regulators or switching operations of less important system elements can all be categorized as small disturbances that occur often during normal system operation. The power system operators and automatic control mechanisms are concerned with regulating the power generation and the power flow inside the system without loss of its steady-state stability.

\subsection{System model}

Consider a power system which consist of $N$ buses of which the last $N-n$ are load buses while the first $n$ are generator buses where there may also be loads.
Each generator is represented by a constant voltage
behind a transient reactance. Furthermore, the loads are represented as constant, passive impedances. This enables us to eliminate all but the generator buses by applying Kron reduction~\cite{dorfler2013kron}.
The full admittance matrix $\textbf{Y}_{bus}$ with dimensions $N \times N$ is thus reduced to a matrix with dimensions $n \times n$. We denote this reduced matrix as $\textbf{Y}_g$. The entries in the reduced matrix $Y_{ik} = G_{ik}+ \jmath B_{ik}$ consist of both a real and imaginary part, denoted by $G_{ik}$ and $B_{ik}$ respectively. 

The swing equation, which governs the mechanical dynamics of the $i$th generator is
\begin{equation}
\label{eq:swing}
M_i\ddot{\delta}_i+D_i\dot{\delta}_i = P_{mi}-P_{ei}
\end{equation}
where $\delta_i$ is the rotor angle, $M_i$ is the rotor inertia constant, $D_i$ is the mechanical damping constant, $P_{mi}$ is the mechanical power, and $P_{ei}$ is the electrical power.
The electrical power at the $i$th generator is given by
\begin{equation}
\label{eq:elect_power}
P_{ei}( \boldsymbol \delta ) = \mbox{Re}  \left[  S_i \right] = \mbox{Re} \left[ V_i I_i^* \right]
\end{equation}
where $V_i$ and $I_i$ are the voltage and current at the $i$th generator.
The mechanical power for the $i$th generator for purposes of steady-state stability analysis is frequently taken as constant $P_{mi} = V_i^2 G_i $, neglecting the effects of the turbine governor~\cite{zimmerman2011matpower}.

It is well known that traditional automatic generation control is implemented locally via a feedback control loop. In an interconnected system these control loops have no information where a power imbalance occurs. 
This could lead to undesirable control actions, possibly overloading certain transmission lines unless manual centralized preventive control action is not applied.

We argue that a change of mechanical output power should be uniformly distributed around the system, that is at power stations evenly located around the system.
There are two reasons that support this claim. 
Firstly, any additional electrical power will come from a variety of locations thus minimizing the risk of overloading some transmission lines. This is important since if certain transmission lines are overloaded that may enable the disturbance to spread.
Secondly, the concentration of mechanical power regulation to a small area poses severe security risks.

To address this problem and in terms to increase the robustness of the system, we propose a control strategy based on a communication infrastructure. 
The proposed strategy utilizes information concerning the relevant phases at neighboring generators that is made available using the communication infrastructure.
This type of control enables a given generator to react to the occurrence of a disturbance in the network even before its effects are manifested as a change in its own relevant phase. This reduces reaction times and allows for a more coordinated action following a disturbance.
Additionally, the control actions can prevent generators with small capacities from getting quickly overloaded by spreading the effects of the disturbance via the communication network to its neighboring generators.

The control can be formally written by adding an extra term in the expression for mechanical power. The notation $\sim$ indicates two generators which are connected via a communication link, and $h_{ik}$ is a negative constant.
\begin{equation}
\label{eq:mech_pow_with_control}
P_{mi}( \boldsymbol \delta ) = P_{mi}^{const} + \sum_{k \sim i} h_{ik} ( \delta_i - \delta_k )
\end{equation}
By substituting (\ref{eq:elect_power}) and (\ref{eq:mech_pow_with_control}) into (\ref{eq:swing}) and writing  $\omega_i=\dot{\delta}_i$, we have:   
\begin{eqnarray}
\label{eq:dynamics_control}
  \dot \delta_i  &=& \omega_i - \omega_s \nonumber \\
  \dot \omega_i  &=& \frac{1}{M_i}  [- D_i(\omega_i - \omega_s)]  + \frac{1}{M_i} \left( P_{mi}^{const}+ \sum_{k \sim i} h_{ik}\delta_{ik} \right) \nonumber \\
  & & + \frac{1}{M_i} \left[ \sum_{k=1}^n  \left( C_{ik} \sin \delta_{ik} + D_{ik} \cos \delta_{ik}  \right) \right], 
\end{eqnarray}
where $\delta_{ik} = \delta_i-\delta_k$, $C_{ik} = |V_i||V_k| B_{ik}$, $D_{ik} = |V_i||V_k| G_{ik}$ and $\omega_s$ represents the angular velocity of a rotating reference frame. Note that the eqs.~(\ref{eq:dynamics_control}) are written with respect to an arbitrary synchronous reference frame.

\subsection{Steady-state stability}

In the presence of small disturbances the equations describing the power system dynamics may be linearized for purposes of performing analytic analysis. Since the linearization is performed in the vicinity of a stable operating point $(\boldsymbol\delta_s,\boldsymbol\omega_s)$, the system coordinates are shifted in order to make the stable equilibrium point the origin. This is done with the following transformation $\Delta  \boldsymbol\delta =  \boldsymbol\delta -  \boldsymbol\delta_s $ and $\Delta  \boldsymbol\omega =  \boldsymbol\omega- \boldsymbol\omega_s$. Assuming the power system model and the control system described previously the linearization of Eqs.~(\ref{eq:dynamics_control}) is
\begin{equation}
\label{eq:linearization_basic}
  \kbordermatrix{%
  & \\
        & \Delta \dot{ \boldsymbol\delta } \\
        & \Delta \dot{ \boldsymbol\omega }
  }
  =
  \kbordermatrix{
  & n & n \\
 n & \textbf{0} & \textbf{1} \\
 n & \textbf{T}+\textbf{K} & \textbf{P} \\
  }
   \kbordermatrix{%
  & \\
        & \Delta \boldsymbol\delta  \\
        & \Delta \boldsymbol\omega
  }
\end{equation}
Note that the Jacobian matrix is given in expanded form for better illustration of its different components. In eq.~(\ref{eq:linearization_basic}) $\textbf{1}$ is an $n \times n$ matrix whose diagonal entries are ones, $\textbf{P}$ is an $n \times n$  diagonal matrix containing the elements $- D_i / M_i$. The elements of $\textbf{T}$ are
\begin{eqnarray}
T_{i,k \neq i} &=& \frac{1}{M_i} \left( D_{ik} \sin \hat \delta_{ik} - C_{ik} \cos \hat \delta_{ik}  \right) \nonumber \\
T_{i,i} &=& \frac{1}{M_i}  \sum_{k=1,k \neq i}^n \left( C_{ik} \cos \hat \delta_{ik}  - D_{ik} \sin \hat \delta_{ik} \right) \nonumber
\end{eqnarray}
The notation $\hat \delta_{ik}$ is used to indicate that the expression is evaluated at the stable equilibrium point $(\boldsymbol\delta_s,\boldsymbol\omega_s)$.
The matrix $\textbf{K}$ is defined as
$K_{i,k \neq i} = \frac{h_i}{M_i}$  if $i \sim k$, 
$K_{i,k \neq i} = 0$ otherwise, and 
$K_{i,i} = -\frac{1}{M_i} \sum_{k \sim i} h_i$.

It is well-known that a non-linear system is (locally) asymptotically stable if its linear approximation is asymptotically stable~\cite{sauer1990power}.  Let $\alpha_k$ be the real part of the $k$th eigenvalue of the Jacobian matrix and ${\alpha_{max} = \max_k \alpha_k, \forall k \in \lbrace1, \ldots ,2n\rbrace}$. The stability condition becomes $\alpha_{max} < 0$. If this condition is satisfied then the speed of convergence \textit{i.~e.} the time for the system to restore stable operation is dominantly determined by the value of $\alpha_{max}$.
Additionally, $\alpha_{max}$ can also be used as an estimate of the overall robustness of a dynamical system~\cite{doyle1982analysis,becker1994robust,chen1999robustness}. In the rest of the paper we will use $\alpha_{max}$ as the main stability indicator for power systems.

\subsection{Problem statement} 

We formulate the problem as follows: \emph{Given a budget $B$ choose the best links in terms of steady--state stability without exceeding that budget.}
For simplicity we assume that all links have the same unit cost i.e. exactly $B$ links can be installed. 
This is an NP-Complete problem and can not be solved by exhaustive search as the number of possible configurations increases exponentially with the number of generators. Thus a suboptimal algorithm must be used.
We develop a simple algorithm to determine which $B$ links should be added to the communication network. The algorithm consists of $B$ iterations, where in each iteration a single link is chosen and added to the communication network. 

Let us denote with $A_i$ the set that contains all links added after the $i$th iteration, with $A_0 = \emptyset $. The gain from installing the links in $A_i$ is
\begin{equation}
{R(A_i) = \alpha_{max}(A_0)-\alpha_{max}(A_i)}
\label{eq:gain}
\end{equation}
where $\alpha_{max}(A_i)$ is the largest real part of all eigenvalues for the Jacobian matrix of the power system with all the communication links in $A_i$ installed. The benefit of adding an arbitrary link $l$ is equal to the difference in the gains for $A_i$ and $A_i \cup l$, \textit{i.e.}
\begin{equation}
\label{eq:marginal_gain}
g_l(A_i) =  R(A_i)-R(A_i \cup l)
\end{equation} 
We refer to $g_l(A_i)$ as the marginal gain for link $l$ with respect to $A_i$.
In the $i$th iteration the link $l_{best}$ that has the maximal marginal gain  is added to $A_i$,~\textit{i.e.}
\begin{eqnarray}
A_{i+1} &=& A_i \cup l_{best} \textrm{ , where} \nonumber \\
l_{best} &=& \max_l g_l(A_i) \hspace{1mm} , \hspace{1mm} \forall \hspace{1mm} l \in S \setminus A, 
\label{eq:maximize}
\end{eqnarray}
where $S$ denotes the set of all possible communication links. After $B$ iterations the solution $A_{K}$ is obtained.

\section{Results}
\label{results}

We tested the proposed control strategy using two test power systems: the IEEE New England~\cite{llamas1995clarifications} and the IEEE RTS 96~\cite{su1997fast,bakirtzis2002optimal}. Overall, the results presented below show that the introduction of a communication infrastructure leads to a decreases in the value of $\alpha_{max}$ indicating increased stability. For both systems a budget $B=15$ and unit link cost was assumed. Under these constraints the best 15 communication links were iteratively added and their effect on the stability of the power systems was observed. Note that, our control scheme requires a control coefficient for each link added in this manner.
Unfortunately, because of the large number of control coefficients and the range of values they can take, testing all possible control configurations is unfeasible.
In order to reduce the complexity of the testing procedure we limit the control coefficients to a common value $h_{ik} = -1, \forall \hspace{1mm} i,k \in \lbrace 1,\ldots ,n \rbrace$. 
The results presented here are thus simply a proof of concept and do not represent the best possible solution which would require further optimization of the control coefficients.

\subsection{New England power system}

The IEEE New England power system is widely used in power system studies. It consist of 10 generators, 17 loads and 39 buses all connected by 47 transmission lines. 
Details of the effect the control strategy has on $\alpha_{max}$ and the marginal gains from each link can be found in Table~\ref{tab:detailsNewEgnalnd}. The first row represents the power system as is without communication links, and serves as a basis for estimating the benefit of the communication infrastructure.

\begin{table}[!htbp]
\renewcommand{\arraystretch}{1.3}
\caption{\hspace{-4.5 mm} Communication links added in each iteration for the IEEE New England power system, their marginal gains and values for $\alpha_{max}$.}
\label{tab:detailsNewEgnalnd}
\centering
\begin{tabular}{ | c | c | c | c | }
  \hline
  	& & &    \\[-8pt]
  Iteration & Generators & $\alpha_{max} \cdot 10^{-2}$ &  $g_l(A_i) \cdot 10^{-4}$
  \\
  \hline
  	& & &    \\[-9pt]
0  &   N/A       & -0.1899 & 0.0000 \\
1  &   1  --   9 & -0.1963 & 0.6360 \\
2  &   1  --   3 & -0.2014 & 0.5195 \\
3  &   1  --   2 & -0.2069 & 0.5483 \\
4  &   1  --   6 & -0.2099 & 0.2921 \\
5  &   1  --  10 & -0.2128 & 0.2995 \\
6  &   1  --   8 & -0.2149 & 0.2023 \\
7  &   1  --   7 & -0.2167 & 0.1848 \\
8  &   8  --   9 & -0.2182 & 0.1499 \\
9  &   3  --   8 & -0.2191 & 0.0929 \\
10 &   2  --   8 & -0.2199 & 0.0763 \\
11 &   9  --  10 & -0.2204 & 0.0444 \\
12 &   3  --  10 & -0.2206 & 0.0268 \\
13 &   2  --  10 & -0.2208 & 0.0185 \\
14 &   6  --   7 & -0.2208 & 0.0034 \\
15 &   8  --  10 & -0.2209 & 0.0027 \\
  \hline	
\end{tabular}
\end{table}

\begin{figure}[!htbp]
\centering
\includegraphics[scale=.9]{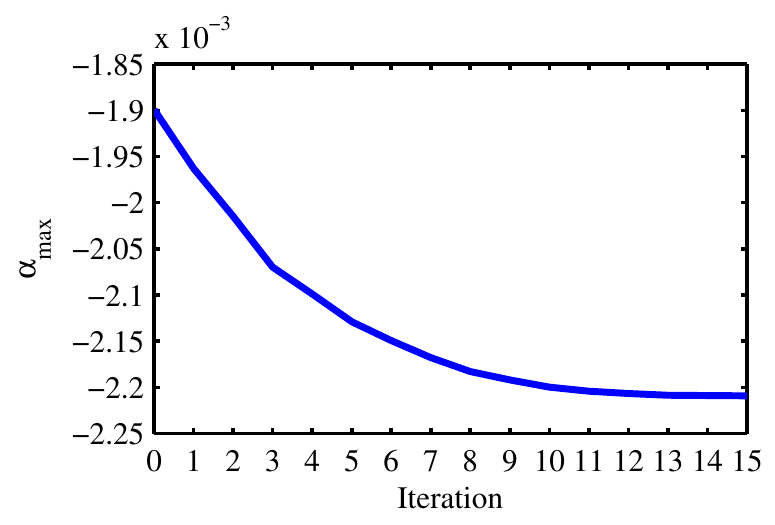}
\caption{IEEE New England: $\alpha_{max}$ decreases as new links are added.}
\label{fig:commnetwork}
\end{figure}

Table~\ref{tab:detailsNewEgnalnd} shows two key trends. Firstly, the value for $\alpha_{max}$ continually decreases with each added link. 
The original system has $\alpha_{max}(0) = -0.1899 \cdot 10^{-2}$, a value that decreases to $\alpha_{max}(15) = -0.2209 \cdot 10^{-2}$, after all 15 links have been added.
Secondly, there is a reduction in the marginal gain. For example, the gain for the 1st communication link is $0.636 \cdot 10^{-4}$, dropping down to $0.5195 \cdot 10^{-4}$ for the second link which is a decrease of 18 percent. The $15$th link only has a marginal gain of $0.27 \cdot 10^{-6}$, which is a 200 times difference compared to marginal gain for the first link. 
Therefore, the stability of the power system  is improved by adding new links, but this effect diminishes as more links are added (Fig.~\ref{fig:commnetwork}). The decrease in $\alpha_{max}$ becomes almost negligible after the $10$th link. 

\begin{figure}[!htbp]
\centering
\includegraphics[scale=0.3]{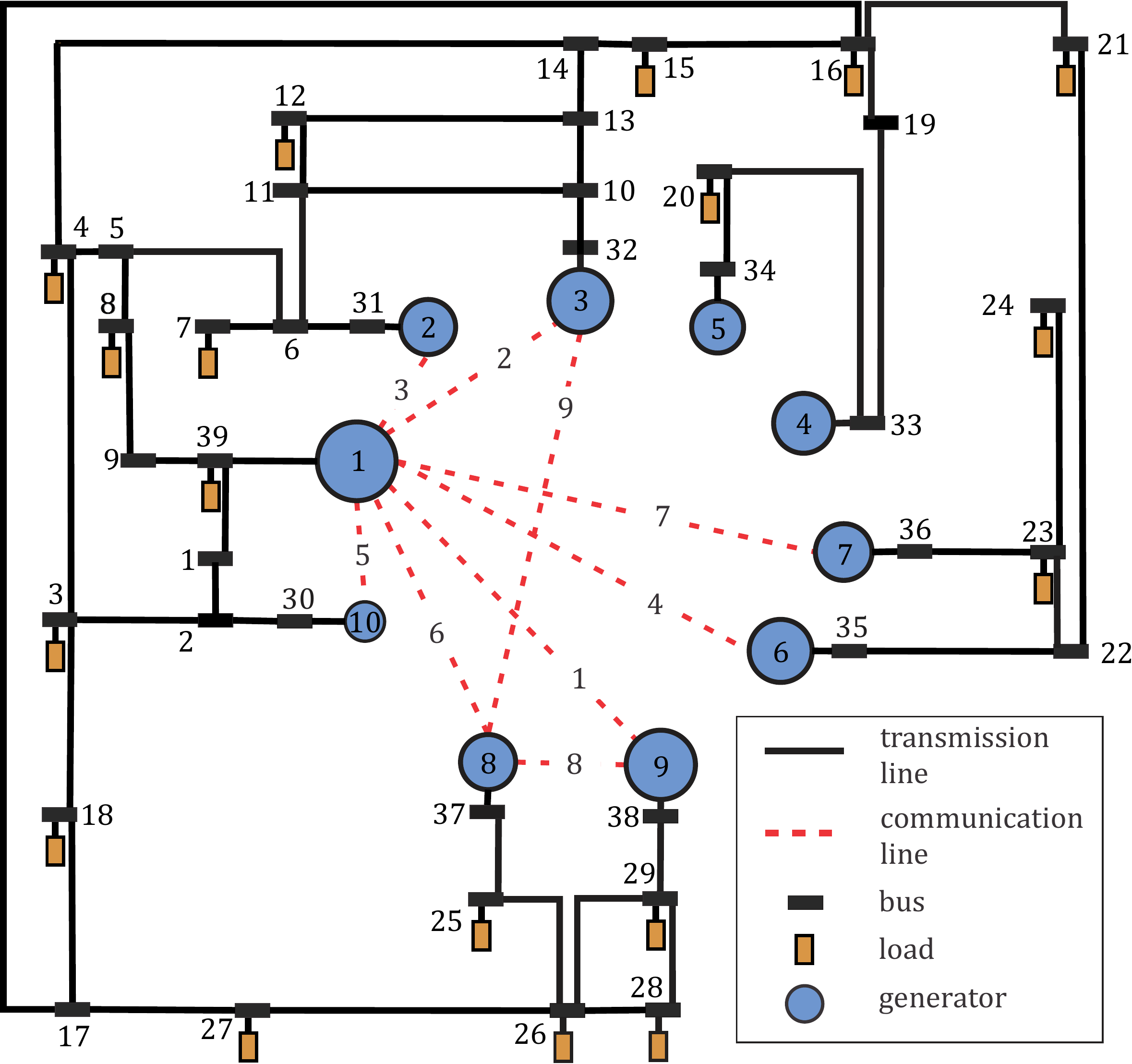}
\caption{Transmission and control topologies for the IEEE New England power system. The numbers on the communication links denote the iterations in which they were added.}
\label{fig:new_england_comm_network}
\end{figure}

The IEEE New England power system with the transmission and control topologies is displayed in Fig.~\ref{fig:new_england_comm_network}. In order to retain distinctness of the presented information only communication links up to the $9$th iteration are shown.
In Fig.~\ref{fig:new_england_comm_network} the generators are depicted as circles with different sizes. The size of the circles is proportional to the electrical power output of the corresponding  generator. The largest and smallest circles, represent the $1$st generator with $P_{e1} = 1000$ MW and the $10$th generator with $P_{e10} = 250$  MW, respectively. The electrical power output of the other machines lies between $500$ and $850$ MW.

The algorithm for constructing the communication network seems to prefer installing links that connect large generators, for example the first link connects the two largest generators. Furthermore, the first $7$ links, from a total of $15$, also include the largest generator. Although, there is a general tendency to connect large generators among themselves no strict pattern can be deduced by inspecting the choice of communication links. This is confirmed by the frequent presence of the smallest generator $10$, which is included in a total of $6$ communication links (Table~\ref{tab:detailsNewEgnalnd}).

\subsection{IEEE RTS 96 system}

In addition to the IEEE New England the control scheme was also tested on the single area IEEE RTS 96 power system. This system consist of $35$ generators, which is considerably more then the $10$ generators in the IEEE New England.
The choice of communication links, their benefit and marginal gain is given in Table~\ref{tab:detailsRTS96}, where the first row corresponds to the original power system without the communication infrastructure.

\begin{figure}[!htbp]
\centering
\includegraphics[scale=.9]{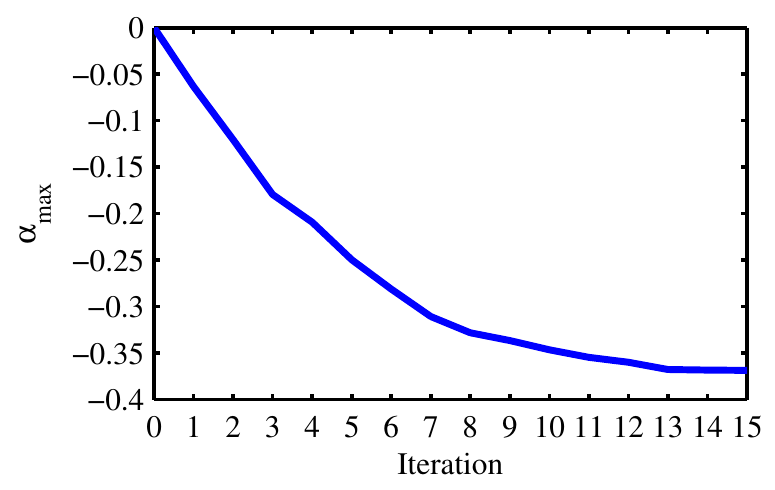}
\caption{IEEE RTS 96: $\alpha_{max}$ decreases as new links are added.}
\label{fig:comm_network_rts_96}
\end{figure}

These results are similar to the ones for the IEEE New England power system in the sense that $\alpha_{max}$ decreases with each iteration, indicating increased stability. 
A comparison between the values for $\alpha_{max}$ for the network without communication infrastructure and after all the links have been added shows a significant improvement in the overall stability; for example 
$\frac{\alpha_{max}(i=15)}{\alpha_{max}(i=0)}=6345$. 
The benefit from adding new links decreases as more links are added (Fig.~\ref{fig:comm_network_rts_96}), which is similar to the results for the New England power system.

\begin{table}[!htbp]
\renewcommand{\arraystretch}{1.3}
\caption{\hspace{-4.5 mm} Communication links added in each iteration for the IEEE RTS 96 power system, their marginal gains and the effects on $\alpha_{max}$.}
\label{tab:detailsRTS96}
\centering
\begin{tabular}{ | c | c | c | c | }
  \hline
  	& & &    \\[-8pt]
  Iteration & Generators & $\alpha_{max}$ & \hspace{2 pt} $ g_l(A_i) \cdot 10^{-1}$ \hspace{2 pt} \\[2pt]
  \hline
  	& & &    \\[-9pt]
0  &   N/A         & $-7 \cdot 10^{-5}$ & 0.0000 \\
1  &   30  --   33 & -0.0635 & 0.6338 \\
2  &   6  --   24  & -0.1206 & 0.5719 \\
3  &   6  --   31  & -0.1794 & 0.5874 \\
4  &   4  --   22  & -0.2094 & 0.2996 \\
5  &   19  --  22  & -0.2499 & 0.4056 \\
6  &   19  --  28  & -0.2813 & 0.3139 \\
7  &   5  --   24  & -0.3108 & 0.2944 \\
8  &   20  --   22 & -0.3282 & 0.1742 \\
9  &   6  --   17  & -0.3366 & 0.0837 \\
10 &   2  --   25  & -0.3466 & 0.1005 \\
11 &   1  --  5    & -0.3547 & 0.0810 \\
12 &   20  --  21  & -0.3601 & 0.0542 \\
13 &   16  --  19  & -0.3677 & 0.0752 \\
14 &   25  --   31 & -0.3685 & 0.0081 \\
15 &   11  --  18  & -0.3688 & 0.0032 \\
  \hline	
\end{tabular}
\end{table}

\section{Conclusions and future work}
\label{concl}

In this paper we have suggested a strategy for improved steady-state stability of power grids by adding communication links to the power grid. Choosing the best links is done by a simple algorithm that maximizes stability.
The resulting communication network topology differs significantly form the transmission grid topology, showing how complex the steady-state control for power systems is. By obtaining a thousandfold decrease for $\alpha_{max}$ we show that the control strategy provides dramatic improvement of the system stability and robustness.
Although the control strategy presented here provides satisfactory results, it leaves much room for improvement. For example, the requirements for PMUs to be installed only at the generator buses can be relaxed to allow installation at major load buses or even arbitrary locations on the power grid. This however will require a different power model to be used, one that preserves the structure of the original power system.
%
%
Another possible research direction that is essential to determine the control strategy potential for practical application is the utilization of more detailed system models both on the generation and load side.
Generators can be more accurately represented by fifth or sixth order model that incorporate the effects of automatic generation control systems together with the influence of damper windings, armature voltage etc.
On the other hand, the loads can be modeled as being voltage and frequency dependent instead of the passive constant impedance model used currently.
%
%
All these questions will be a subject of our future research.

\end{document}